\documentclass[twocolumn,showpacs,prl,superscriptaddress]{revtex4}
\usepackage{graphicx}
\usepackage[usenames]{color}
\usepackage{amssymb}
\usepackage{amsmath}
\usepackage{amsfonts}
\usepackage{mathrsfs}

\renewcommand{\epsilon}{\varepsilon}
\newcommand{\figurewidth}{0.46\textwidth}

\begin{document}
\title{Chain conformation of ring polymers under a cylindrical nanochannel confinement}

\author{Junfang Sheng}
\affiliation{CAS Key Laboratory of Soft Matter Chemistry, Department of Polymer Science and Engineering, University of Science and Technology of China, Hefei, Anhui Province 230026, P. R. China}

\author{Kaifu Luo}
\altaffiliation[]{Author to whom the correspondence should be addressed}
\email{kluo@ustc.edu.cn}
\affiliation{CAS Key Laboratory of Soft Matter Chemistry, Department of Polymer Science and Engineering, University of Science and Technology of China, Hefei, Anhui Province 230026, P. R. China}

\date{\today}

\begin{abstract}

We investigate the chain conformation of ring polymers confined to a cylindrical nanochannel using both theoretical analysis and three dimensional Langevin dynamics simulations.
We predict that the longitudinal size of a ring polymer scales with the chain length and the diameter of the channel in the same manner as that for linear chains based on scaling analysis and Flory-type theory. Moreover, Flory-type theory also gives the ratio of the longitudinal sizes for a ring polymer and a linear chain with identical chain length. These theoretical predictions are confirmed by numerical simulations. Finally, our simulation results show that this ratio first decreases and then saturates with increasing the chain stiffness, which has interpreted the discrepancy in experiments. Our results have biological significance.

\end{abstract}


\maketitle

\textit{Introduction}.
The ring polymer, also called cyclic polymer, is an important member of the polymer family. In biopolymer science, there exist circular DNA, cyclic peptides and cyclic oligosaccharides and polysaccharides \cite{Cyclic}.

The statistical mechanical properties of polymer are influenced by the ring closure. One of the fundamental properties for ring polymers is the chain sizes in solution. In a $\theta$ solvent, based on Gaussian approximation Zimm and Stockmayer \cite{Zimm} have shown the radius of gyration of ideal ring polymers, $R_{g,r}\sim N^{\nu}$ with $\nu=0.5$, which scales with the chain length $N$ in the same manner as that of ideal linear polymers, $R_{g,l}$, and further found the ratio $R_{g,r}/R_{g,l}=1/\sqrt{2}=0.707$ for ring polymers and linear chains of the same chain length. Under a good solvent conditions, based on a renormalized two-parameter theory Douglas and Freed \cite{Freed} have demonstrated $R_{g,r}\sim N^{\nu}$ with $\nu=0.6$, where the scaling exponent is also the same as that for a linear polymer. In addition, they have also found the ratio $R_{g,r}/R_{g,l}=0.718$, which is larger than that for ideal chains. The scaling exponent $\nu=0.6$ is in agreement with other theoretical results \cite{Cloizeaux,Deutsch,Grosberg} and experiments \cite{Matsushita}, but slightly larger values of $R_{g,r}/R_{g,l}$ are observed in simulations and experiments \cite{Cyclic}.

Motivated by the fundamental relevance in polymer physics and many biological processes, such as DNA packaging inside the phage capsid \cite{Smith}, polymer translocation through nanopores \cite{Luo1,Luo2} and viruses injecting their DNA into a host cell \cite{Miller}, the properties of a polymer confined in a nanochannel have been the subject of extensive studies \cite{Brochard,Kremer,Sotta,Cifra,Gong,Klushin,Yang,Jun,Dorier,Marenduzzo,Matthews,Obukhov,Reisner,Persson,Witz}. %
Based on the blob picture \cite {Gennes}, the conformational properties of a linear self-avoiding polymer chain confined in a slit or in a cylindrical nanochannel are relatively well understood \cite {Daoud,Gennes,Milchev10,Milchev11}.
Unlike its linear polymer counterpart, the physics of confined ring polymers is still not clear. Only recently, few studies have addressed semiflexible ring polymers. For semiflexible ring polymers in weak spherical confinement, the internal structure of the chain shows buckling and a conformational transition to a figure eight form \cite{Frey}. For the conformational properties of a semiflexible ring polymer confined to different geometrical constraints, Fritsche and Heermann \cite{Heermann} found an important role of the geometry of confinement in shaping the spatial organization of polymers.

However, even for flexible ring polymers confined to a cylindrical nanochannel, the behavior of the chain remains unclear. The basic questions associated with flexible ring polymers confined to a cylindrical nanochannel are the following: (a) What's the scaling behavior of the longitudinal size with the chain length and the channel diameter? (b) What's the ratio of the longitudinal sizes for a ring polymer and the linear chain of the same length? The second question deals with the numerical prefactors of the scaling behaviors, which may change our understanding about polymers. On the one hand, the numerical prefactors is very important for understanding the biopolymer dynamics, such as the chain ejection out of the nanochannel where the confinement induced driving force is related to the numerical prefactor. On the other hand, knowing numerical prefactors can help us deduce the chain conformation of ring polymers confined to a cylindrical nanochannel by comparison with the wellknown chain conformation of the confined linear chain.
To this end, in this work we examine the properties of ring polymers confined in nanochannel and their differences from the linear chains.

\textit{Model and methods}.
In simulations, the polymer is coarse-grained as bead-spring chain of Lennard-Jones (LJ) particles with the Finite Extension Nonlinear Elastic (FENE) potential. We use a short range repulsive LJ potential to model the excluded volume interaction between beads: $U_{LJ} (r)=4\epsilon [{(\frac{\sigma}{r})}^{12}-{(\frac{\sigma} {r})}^6]+\epsilon$ for $r\le 2^{1/6}\sigma$ and 0 for $r>2^{1/6}\sigma$. Here, $\sigma$ is the diameter of a bead, and $\epsilon$ is the depth of the potential. The connectivity between neighboring beads is modeled by a FENE spring with $U_{FENE}(r)=-\frac{1}{2}kR_0^2\ln(1-r^2/R_0^2)$, where $r$ is the distance between consecutive beads, $k$ is the spring constant and $R_0$ is the maximum allowed separation between connected beads.

The nanochannel is described as stationary particles within distance $\sigma$ from one another which interact with the beads by the repulsive Lennard-Jones potential. The nanochannel particle positions are not changed in the simulations.
Each bead is subjected to conservative, frictional, and random forces in the Langevin dynamics simulation \cite{Allen}:  $m{\bf \ddot {r}}_i =-{\bf \nabla}({U}_{LJ}+{U}_{FENE})-\xi {\bf v}_i+{\bf F}_i^R$. Here $m$ is the bead's mass, $\xi$ is the friction coefficient, ${\bf v}_i$ is the bead's velocity, and ${\bf F}_i^R$ is the random force which satisfies the fluctuation-dissipation theorem. In the simulation, we use the LJ parameters $\epsilon$,
$\sigma$, and $m$ to fix the system energy, length and mass units, respectively. Thus, we have the corresponding time scale $t_{LJ}=(m\sigma^2/\epsilon)^{1/2}$ and force scale $\epsilon/\sigma$, which are of the order of ps and pN, respectively. The dimensionless parameters in the model are then chosen to be $R_0=1.5$, $k=15$, $\xi=0.7$.

In our model, each bead corresponds to a Kuhn length (twice of the persistence length) of a polymer. For a single-stranded DNA (ssDNA), the persistence length of the ssDNA is sequence and solvent dependent and varies in a wide range \cite{Tinland,Smith2}. Here, we choose the value of $\sigma \sim 2.8$ nm for a ssDNA containing approximately four nucleotide bases \cite{Smith2}. The average mass of a base in DNA is about 312 amu, leading to the bead mass $m \approx 1248$ amu. We set $k_{B}T=1.2\epsilon$, and thus the interaction strength $\epsilon$ is $3.39 \times 10^{-21}$ J at actual temperature 295 K. This leads to a time scale of 69.2 ps and a force scale of 1.2 pN.
The Langevin equation is then integrated in time by a method described by Ermak and Buckholz \cite{Ermak}.

\textit{Results and discussion}.
One of the central properties of a polymer confined in a cylindrical nanochannel is the longitudinal size of the polymer.
Consider a flexible polymer confined to a cylindrical nanochannel of diameter $D$ which is less than $R_g$, the radius of gyration of the chain. The Response of the polymer to confinements is primarily dictated by the relative value of $D$ with respect to the chain persistence length. Depending on whether $D$ is larger (de Gennes regime \cite{Gennes}) or smaller (Odijk regime \cite{Odijk}) than the chain persistence length, different scaling behaviors of the longitudinal size of the chain, $R_{\parallel}$, as a function of $D$ were predicted in the pioneering theoretical studies by de Gennes \cite{Gennes} and Odijk \cite{Odijk}, respectively. In the Odijk regime, the physics is decimated not by excluded volume but by the interplay of confinement and intrinsic polymer elasticity. To consider Odijk regime, we need to take into account the chain stiffness in the model. In this work, we only consider the de Gennes regime, namely $\sigma \ll D\ll R_g$ with $\sigma$ being the length of a segment of the chain.

For a single polymer chain confined in a cylindrical nanochannel of diameter $D$, there are two characteristic lengths, $D$ and $R_g$.
Based on the scaling analysis \cite{Gennes}, the length occupied by the chain in the nanochannel, $R_{\parallel}$ has the scaling form
\begin{equation}
R_{\parallel}=R_g\phi(\frac{R_g}{D}),
\label{eq1}
\end{equation}
where $\phi(x)\sim x^m$ is a dimensionless scaling function. In the de Gennes regime, the chain becomes a one-dimensional chain, noting to do with the chain topology (ring polymers or linear chains). Therefore, for both ring polymers and linear chains, $R_{\parallel}\sim N$. As a sequence, $m= 1/\nu -1$ with $\nu=0.588$ is the flory exponent in three dimensions \cite {Gennes}. According to Eq. (\ref{eq1}), we have
\begin{equation}
R_{\parallel,r}\sim R_{\parallel,l} \sim N\sigma (\frac{\sigma}{D})^{1/\nu -1},
\label{eq2}
\end{equation}
where, $R_{\parallel,r}$ and $R_{\parallel,l}$ are the lengths occupied by the chain in the nanochannel for ring polymers and linear chains, respectively.
Although above scaling analysis clearly provides the physics for $R_{\parallel}$, we still cannot obtain the ratio $\frac {R_{\parallel,r}} {R_{\parallel,l}}$. To this end, we use Flory-like theory for both linear chains and ring polymers.

According to the blob picture, for a linear chian confined in an infinitely long nanochannel of diameter $D$, the chain will extend along the channel axis forming a string of blobs of diameter $\xi_{b,l}=D$. On length scales smaller than the blob size, the effects of the boundaries are weak and the subchains in the blob follow excluded volume statistics. Namely, for each blob $\xi_{b,l}=Cg_l^\nu\sigma$, where $g_l$ is the number of monomers in a blob. The prefactor $C$ is a constant, which depends on the temperature and the solvent. Thus, each blob contains $g_l=(\frac{\xi_{b,l}}{C\sigma})^{\frac{1}{\nu}}$ monomers, and the number of blobs is $n_b=N/g_l=N(\frac{C\sigma}{\xi_{b,l}})^{\frac{1}{\nu}}$. Considering the entropic elasticity of the polymer and the excluded volume repulsion between the blobs. The free energy of a linear chian confined in a nanochannel is given by \cite{Jun2}
\begin{equation}
\frac{\mathcal{F}_{linear}}{k_BT}=A\frac{R_{\parallel,l}^2}{D^2(N/g_l)}+B\frac{D(N/g_l)^2}{R_{\parallel,l}},
\label{eq3}
\end{equation}
where $A$ and $B$ are constants, the first term is the elastic free energy, and the second term represents excluded volume interactions.
Minimizing the free energy with respect to $R_{\parallel,l}$ leads to the extension $R_{\parallel,l}$ of polymer trapped in the nanochannel, we have
\begin{equation}
R_{\parallel,l}=(\frac{B}{2A})^{1/3}D(N/g_l)=(\frac{B}{2A})^{1/3}ND^{1-1/\nu}(C\sigma)^{1/\nu}.
\label{eq4}
\end{equation}
Correspondingly, the free energy at $R_{\parallel,l}$ is
\begin{equation}
\frac{\mathcal{F}_{linear}}{k_BT}=\frac{3}{2^{2/3}}A^{1/3}B^{2/3}ND^{-1/\nu}(C\sigma)^{1/\nu}.
\label{eqF1}
\end{equation}

For a ring polymer confined in an infinitely long nanochannel of diameter $D$, there are always two folded subchains inside the channel, which forms two string of blobs of size $\xi_{b,r}$. In each blob, $\xi_{b,r}=Cg_r^\nu\sigma$ with $g_r$ being the number of monomers in a blob. The free energy due to excluded volume interactions is $B\frac{\xi_{b,r}^3}{D^2}\frac{(N/g_r)^2}{R_{\parallel,r}}$. In contrast to the linear chain, the ring polymer can been considered as two subchains of length $N/2$, and thus the elastic free energy is $2A\frac{R_{\parallel,r}^2}{\xi_{b,r}^2(N/2g_r)}$. Therefore, the total free energy of a ring polymer confined in a nanochannel is
\begin{equation}
\frac{\mathcal{F}_{ring}}{k_BT}=2A\frac{R_{\parallel,r}^2}{\xi_{b,r}^2(N/2g_r)}+B\frac{\xi_{b,r}^3}{D^2}\frac{(N/g_r)^2}{R_{\parallel,r}}.
\label{eq5}
\end{equation}
The equilibrium size, $R_{\parallel,r}$, at which $\mathcal{F}_{ring}$ is minimized, is
\begin{equation}
R_{\parallel,r}=(\frac{B}{8A}\frac{\xi_{b,r}^5}{D^2})^{1/3}(N/g_r).
\label{eq6}
\end{equation}
Using $\xi_{b,r}=Cg_r^\nu\sigma$ and the Flory exponent $\nu=3/5$, we further obtain
\begin{equation}
R_{\parallel,r}=(\frac{B}{8A})^{1/3}ND^{1-1/\nu}(C\sigma)^{1/\nu}.
\label{eq7}
\end{equation}
Interestingly, \textit{$R_{\parallel,r}$ is independent of the value of $\xi_{b,r}$.}

Obviously, Eqs. (\ref{eq4}) and (\ref{eq7}) derived from the Flory-like theory are in good agreement with the Eq. (\ref{eq2}) from the scaling analysis. Moreover, Eqs. (\ref{eq7}) and (\ref{eq4}) also give the ratio $\frac {R_{\parallel,r}} {R_{\parallel,l}}$ as
\begin{equation}
\frac{R_{\parallel,r}}{R_{\parallel,l}}=(\frac{1}{4})^{1/3}=0.630.
\label{eq8}
\end{equation}
Compared with the value 0.718 for unconfined chains, this ratio is smaller.

However, the free energy of the chain at $R_{\parallel,r}$ is a function of $\xi_{b,r}$, which is
\begin{equation}
\frac{\mathcal{F}_{ring}}{k_BT}=3A^{1/3}B^{2/3}ND^{-4/3}\xi_{b,r}^{4/3-1/\nu} (C\sigma)^{1/\nu}.
\label{eqF2}
\end{equation}
It is difficult to determine $\xi_{b,r}$. Taking into account geometric average, $\pi (\xi_{b,r}/2)^2=0.5\pi(D/2)^2$, one obtains $\xi_{b,r}=\frac{\sqrt{2}}{2}D$. However, the geometric average here is questionable. Statistically, the blob should be spherical because the segments in the blob cannot feel the existence of the wall. If we still use $\xi_{b,r}=\frac{\sqrt{2}}{2}D$, Eq. (\ref{eqF2}) is rewritten as
\begin{equation}
\frac{\mathcal{F}_{ring}}{k_BT}=\frac{3}{2^{2/3-1/2\nu}}A^{1/3}B^{2/3}ND^{-1/\nu} (C\sigma)^{1/\nu}.
\label{eqF3}
\end{equation}

As to the ratio the ratio $\frac {\mathcal{F}_{ring}} {\mathcal{F}_{linear}}$, Eqs. (\ref{eqF3}) and (\ref{eqF1}) give
\begin{equation}
\frac {\mathcal{F}_{ring}} {\mathcal{F}_{linear}}=2^{5/6}=1.782.
\label{eq9}
\end{equation}

For long chains partially confined in the nanochannel, it can escape from the channel by the entropic force. This pulling force can be estimated from the free energy and the longitudinal size, where prefactors of scaling behaviors for the free energy and the longitudinal size are very important.
The ratio of entropic forces for a ring polymer and a linear polymer is
\begin{equation}
\frac {f_{ring}} {f_{linear}}=2^{2/3+1/2\nu}=2\sqrt{2},
\label{eq10}
\end{equation}
where $\nu=3/5$ is used. Even for one string of the long ring polymer, the pulling force acting on it is $\sqrt{2}$ times as that for the long linear chains. This indicates that long ring polymer are easier to escape from the channel than long linear polymers due to smaller blob size in the channel for ring polymers.

\begin{figure}
\includegraphics*[width=\figurewidth]{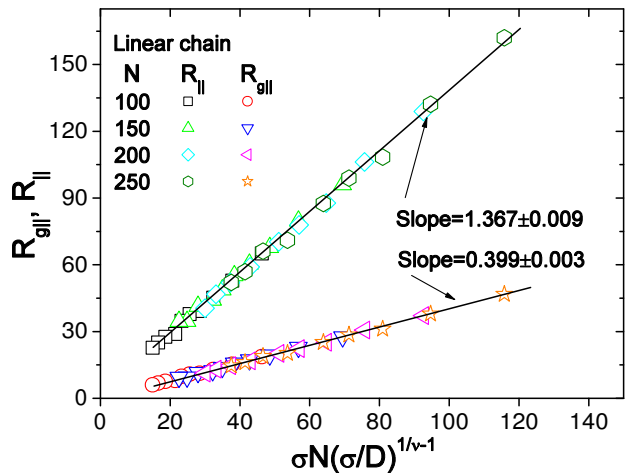}
\caption{Longitudinal size $R_{\parallel}$ and the radius of gyration $R_{g\parallel}$ along the channel axis of flexible self-avoiding linear chains as a function of $\sigma N(\sigma/D)^{\frac{1}{\nu}-1}$. Different polymer lengths $N$ and channel diameters $D$ are used. In the plots, we use $\nu=0.588$. The slopes, which stand for the prefactors of the scaling behaviors, are 1.367 and 0.399 for $R_{\parallel}$ and $R_{g\parallel}$, respectively.
        }
\label{Fig1}
\end{figure}

\begin{figure}
\includegraphics*[width=\figurewidth]{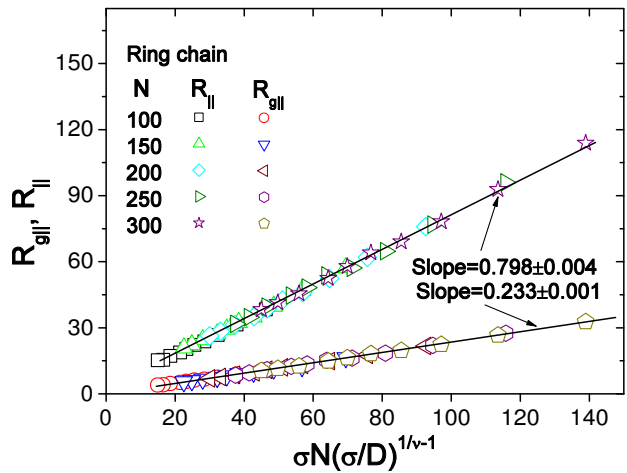}
\caption{Longitudinal size $R_{\parallel}$ and the radius of gyration $R_{g\parallel}$ along the channel axis of flexible self-avoiding ring polymers as a function of $\sigma N(\sigma/D)^{\frac{1}{\nu}-1}$.  Different polymer lengths $N$ and channel diameters $D$ are used. In the plots, we use $\nu=0.588$. The slopes, which stand for the prefactors of the scaling behaviors, are 0.798 and 0.233 for $R_{\parallel}$ and $R_{g\parallel}$, respectively.
        }
\label{Fig2}
\end{figure}

Fig. \ref{Fig1} shows the longitudinal size $R_{\parallel,l}$ and the radius of gyration $R_{g\parallel,l}$ along the channel axis as a function of $\sigma N(\sigma/D)^{\frac{1}{\nu}-1}$ for flexible linear chains. Here, $N$ is the chain length, $\sigma$ is the Kuhn length of the chain, and $\nu=0.588$ is the Flory exponent in three dimensions \cite{Gennes}. As expected, all the data for $R_{\parallel,l}$ and $R_{g\parallel,l}$ collapse onto two different straight lines for different chain length $N$ and channel diameter $D$, respectively. This is in agreement with previous Monte Carlo simulation results \cite{Milchev10}. The slopes, which stand for the prefactors of the scaling behaviors, are $1.367\pm 0.009$ and $0.399\pm 0.003$ for $R_{\parallel,l}$ and $R_{g\parallel,l}$, respectively.

For flexible ring polymers, $R_{\parallel,r}$ and $R_{g\parallel,r}$ as a function of $\sigma N(\sigma/D)^{\frac{1}{\nu}-1}$  for different chain length $N$ and channel diameter $D$ are plotted in Fig. \ref{Fig2}. Interestingly, all the data for $R_{\parallel,r}$ and $R_{g\parallel,r}$ also collapse onto two different straight lines, respectively.
The slopes are $0.798\pm0.004$ and $0.233\pm0.001$ for $R_{\parallel,r}$ and $R_{g\parallel,r}$, respectively.

These results demonstrate that a confined ring polymer behaves like a linear one, although they have different chain topology, in agreement with predictions in Eqs. (\ref{eq2}), (\ref{eq4}) and (\ref{eq7}). Persson \textit{et al.} \cite{Persson} have measured the scaling of circular DNA extension with channel diameter. The scaling exponent they found is in agreement with our prediction in Eq. (\ref{eq7}).

The prefactor of the scaling behavior for the longitudinal size is a very important measure for the static and dynamics of the chain confined to the nanochannel. On the one hand, it is the extension factor of a chain in a nanochannel. Based on slopes of the curves in Figs. \ref{Fig1} and \ref{Fig2}, we have $\frac{R_{{\parallel},r}}{R_{{\parallel},l}}=\frac{0.798}{1.367}\approx 0.584$ and $\frac{R_{g\parallel,r}}{R_{g\parallel,l}}=\frac{0.233}{0.399}\approx 0.584$, which are quite close to the prediction in Eq. (\ref{eq8}), reflecting that the ring polymer has less extension than the linear one with the same chain length due to the ring closure.
At the same time, it demonstrates that the ring polymer with chain length $2N$ is 16.8\% more extended than the linear chain of length $N$, and implies that for two overlapping polymers in a nanochannel each has a larger extension than that in the absence of the other.

Recently, Levy \textit{et al.} \cite{Levy} presented results concerning on partly folded DNA confined in straight nanochannels, where a dyed linear DNA was forced into the channel in a folded configuration (where a portion near the middle of the molecule enters first) by applying an electric field. They observed that the equilibrium extension per unit length for the folded portion was approximately 30\% larger than the equilibrium extension for unfolded molecules. However, using confine spectroscopy, Persson \textit{et al.} \cite{Persson} found that the extension at a given confinement for circular DNA is approximately 5\% less than that for linear DNA for channel dimensions similar to those presented by Levy \textit{et al.} \cite{Levy}. Obviously, there is a discrepancy between these two experiments.

\begin{figure}
\includegraphics*[width=\figurewidth]{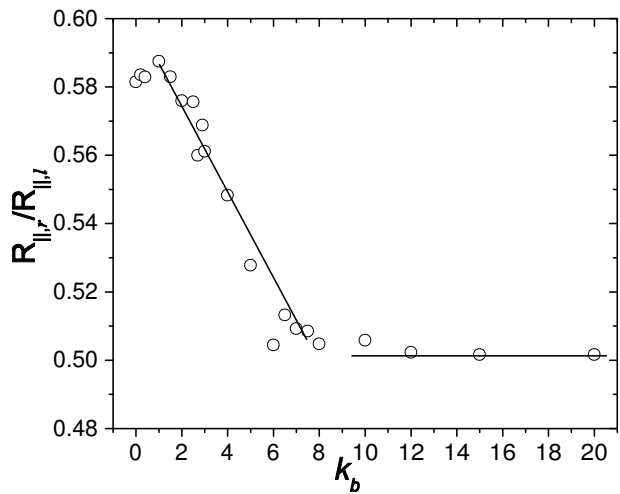}
\caption{Effect of the chain stiffness on $\frac{R_{\parallel,r}}{R_{\parallel,l}}$. Here, $k_b$ is the bending constant. The chain stiffness increases with $k_b$.
        }
\label{Fig3}
\end{figure}

To solve this discrepancy, in the simulation we take into account the chain stiffness by by a harmonic angle potential $U_b(\theta)=k_b(\theta-\pi)^2$, where $\theta$ is the bond angle between two consecutive bonds
and $k_b$ is the bending constant. The chain stiffness increases with $k_b$. Fig. \ref{Fig3} shows that $\frac{R_{{\parallel},r}}{R_{{\parallel},l}}$ first decreases from about 0.58 and then saturates to about 0.5 with increasing the chain stiffness. In above two experiments, different solvents are used which can lead to different stiffness of the DNA molecule. Thus, the discrepancy between these two experiments is from different chain stiffness.


Nature not only imposes geometrical constraints on biopolymers by confinement through cell membrane, the cell nucleus or viral capsid, but also takes advantages of certain underlying chain topologies, such as the ring structure. In fact, \textit{E. coli} has a rod-shaped geometry and its chromosome is not a linear polymer but a circular one.
%
%
Based on Monte Carlo simulations, Jun and Mulder \cite{Jun} have addressed a basic physical issue associated with bacterial chromosome segregation in rod-shaped cell-like geometry and found that two ring polymers segregate more readily than linear ones in confinement. According to our above theoretical analysis and simulation results, for ring polymers confined in a cylindrical nanochannel the blob size is smaller than that for linear polymers. This means that the pulling force for the ring polymer is larger than that for the linear one during the chromosome segregation as shown in Eq. (\ref{eq10}), leading to a faster segregation \cite {Sheng}.


\textit{Conclusions}.
We investigate the chain conformation of ring polymers confined to a cylindrical nanochannel using both theoretical analysis and three dimensional Langevin dynamics simulations.
We predict that the longitudinal size of a ring polymer scales with the chain length and the diameter of the channel in the same manner as that for linear chains based on scaling analysis and Flory-type theory. Moreover, Flory-type theory also gives the ratio of the longitudinal sizes for a ring polymer and a linear chain with identical chain length. These theoretical predictions are confirmed by numerical simulations. Finally, our simulation results show that this ratio first decreases and then saturates with increasing the chain stiffness, which has interpreted the discrepancy in experiments.

Our results should enable a new understanding of the conformation statistics of confined ring biopolymers such as DNA. We believe that this work is important for understanding biological systems with more complexity, such as the behavior of DNA inside phages or the spatial organization of the bacterial nucleoid in \textit{E. coli}.

\begin{acknowledgments}
This work is supported by the National Natural Science Foundation of China (Grant Nos. 21074126, 21174140), the Specialized Research Fund for the Doctoral Program of Higher Education (Grant No. 20103402110032), and the ``Hundred Talents Program'' of Chinese Academy of Science (CAS).
\end{acknowledgments}


\begin{thebibliography}{8}

\bibitem{Cyclic} J. A. Semlyen, \textit{Cyclic Polymers}, 2nd ed. (Springer, Dordrecht, 2000).
\bibitem{Zimm} B. H. Zimm, and W. H. Stockmayer, \textit{J. Chem. Phys.} {\bf 17}, 1301 (1949).
\bibitem{Freed} J. F. Douglas, and K. F. Freed, \textit{Macromolecules} {\bf 17}, 2344 (1984).
\bibitem{Cloizeaux} J. des Cloizeaux, \textit{J. Phys. Lett.} {\bf 42}, 433 (1981).
\bibitem{Deutsch} J. M. Deutsch, \textit{Phys. Rev. E.} {\bf 59}, R2539 (1999).
\bibitem{Grosberg} A. Yu. Grosberg, \textit{Phys. Rev. Lett.} {\bf 85}, 3858 (2000).
\bibitem{Matsushita} A. Takano, Y. Ohta, K. Masuoka, K. Matsubara, T. Nakano, A. Hieno, M. Itakura, K. Takahashi, S. Kinugasa, D. Kawaguchi, Y. Takahashi, and Y. Matsushita, \textit{Macromolecules} {\bf 17}, 2344 (1984).
%
\bibitem{Smith} D. E. Smith, S. J. Tans, S. B. Smith, S. Grimes, D. L. Anderson, and C. Bustamante, \textit{Nature} {\bf 413}, 748 (2001).
\bibitem{Luo1} K. Luo, T. Ala-Nissila, and S. C. Ying,  \textit{J. Chem. Phys.} {\bf 124}, 034714 (2006).
\bibitem{Luo2} K. Luo, T. Ala-Nissila, S. C. Ying, and A. Bhattacharya, \textit{Phys. Rev. Lett.} {\bf 100}, 058101 (2008).
\bibitem{Miller} R. V. Miller, \textit{Sci. Am.} {\bf 278}, 66 (1998).
\bibitem{Brochard} F. Brochard-Wyart, and P. G. de Gennes, \textit{J. Chem. Phys.} {\bf 67}, 52 (1977).
\bibitem{Kremer} K. Kremer, and K. Binder, \textit{J. Chem. Phys.} {\bf 81}, 6381 (1984).
\bibitem{Sotta} P. Sotta, A. Lesne, and J. M. Victor, \textit{J. Chem. Phys.} {\bf 112}, 1565 (2000).
\bibitem{Cifra} P. Cifra, \textit{J. Chem. Phys.} {\bf 131}, 224903 (2009).
\bibitem{Gong} Y. Gong, and Y. Wang, \textit{Macromolecules.} {\bf 35}, 7492 (2002).
\bibitem{Klushin} L. I. Klushin, A. M. Skvortsov, H. P. Hsu, and K. Binder, \textit{Macromolecules.} {\bf 41}, 5890 (2008).
\bibitem{Yang} Y. Yang, T. W. Burkhardt, and G. Gompper, \textit{Phys. Rev. E.} {\bf 76}, 011804 (2007).
\bibitem{Jun} S. Jun and B. Mulder, \textit{Proc. Natl. Acad. Sci. U.S.A.} {\bf 103}, 12388 (2006).
%
\bibitem{Dorier} J. Dorier and A. Stasiak, \textit{Nucleic Acids Res.} {\bf 37}, 6316 (2009).
\bibitem{Marenduzzo} D. Marenduzzo and C. Micheletti, \textit{J. Mol. Biol.} {\bf 330}, 485 (2003).
\bibitem{Matthews} R. Matthews, A. A. Louis, and J. M. Yeomans, \textit{Phys. Rev. Lett.} {\bf 102}, 088101 (2009).
\bibitem{Obukhov} S. P. Obukhov, M. Rubinstein, and T. Duke, \textit{Phys. Rev. Lett.} {\bf 73}, 1263 (1994).
\bibitem{Reisner} W. Reisner, K. J. Morton, R. Riehn, Y. M. Wang, Z. Yu, M. Rosen, J. C. Sturm, S. Y. Chou, E. Frey, and R. H. Austin, \textit{Phys. Rev. Lett.} {\bf 94}, 196101 (2005).
%
\bibitem{Persson} F. Persson, P. Utko, W. Reisner, N. B. Larsen, and A. Kristensen, \textit{Nano Lett.} {\bf 9}, 1382 (2009).
\bibitem{Witz} G. Witz, K. Rechendorff, J. Adamcik, and G. Dietler, \textit{Phys. Rev. Lett.} {\bf 106}, 248301 (2011).
%
\bibitem{Gennes} P. G. de Gennes, \textit{Scaling Concepts in Polymer Physics} (Cornell University Press, Ithaca, NY, 1979).
\bibitem{Daoud} M. Daoud M and P. G. de Gennes, \textit{J. Physique} {\bf 38}, 85 (1977).
\bibitem{Milchev10} A. Milchev, L. Klushin, A. Skvortsov ,and K. Binder, \textit{Macromolecules} {\bf 43}, 6877 (2010).
\bibitem{Milchev11} A. Milchev, \textit{J. Phys.: condens. Matter} {\bf 23}, 103101 (2011).

\bibitem{Frey} K. Ostermeir, K. Alim, and E. Frey, \textit{Phys. Rev. E} {\bf 81}, 061802 (2010); \textit{Soft Matter} {\bf 6}, 3467 (2010)
\bibitem{Heermann} M. Fritsche, and D. W. Heermann, \textit{Soft Matter} {\bf 7}, 6906 (2011).
%
\bibitem{Allen} M. P. Allen and D. J. Tildesley, \textit{Computer Simulation of Liquids} (Oxford University, New York, 1987)
\bibitem{Tinland} B. Tinland, A. Pluen, J. Sturm, and G. Weill, \textit{Macromolecules} {\bf 30}, 5763 (1997).
\bibitem{Smith2} S. B. Smith, Y. Cui, and C. Bustamante, \textit{Science} {\bf 271}, 795 (1996).
\bibitem{Ermak} D. L. Ermak and H. Buckholz, \textit{J. Comput. Phys.} {\bf 35}, 169 (1980).
\bibitem{Odijk} T. Odijk, \textit{Macromolecules} {\bf 16}, 1340 (1983).
\bibitem{Levy} S. L. Levy, J. T. Mannion, J. Cheng, C. H. Reccius and H. G. Craighead, \textit{Nano Lett.} {\bf 8}, 3839 (2008).
\bibitem{Jun2} S. Jun, D. Thirumalai, and B. Y. Ha, \textit{Phys. Rev. Lett.} {\bf 101}, 138101 (2008).
\bibitem{Sheng} J. Sheng and K. Luo, \textit{Soft Matter} {\bf 8}, 369 (2012).
\end{thebibliography}
\end{document}